\documentclass[reprint, prd, superscriptaddress, longbibliography, tightenlines, nofootinbib, eqsecnum, floatfix, notitlepage,twocolumn]{revtex4-1}
\pdfoutput=1

\usepackage[utf8]{inputenc}
\usepackage{euscript}
\usepackage{epsfig}
\usepackage{graphics}
\usepackage{graphicx}
\usepackage{amsmath}
\usepackage{amssymb}
\usepackage{amsfonts}
\usepackage{bm}
\usepackage[usenames,dvipsnames,svgnames,table]{xcolor}
\usepackage{xspace}
\usepackage{times}
\usepackage{appendix}
\usepackage{lipsum}
\usepackage[nolist,nohyperlinks]{acronym}
\usepackage{float}
\usepackage{simplewick}
\usepackage{tabularx}
\usepackage{booktabs}
\usepackage{natbib, ifthen}
\usepackage{hyperref}
\usepackage{comment}
\usepackage{soul}

\newcommand{\vl}{\ensuremath{\boldsymbol{l}}\xspace}

\newcommand{\LM}[0]{{LM}}

\newcommand{\nlzero}{\ensuremath{N_L^{(0)}}\xspace}

\newcommand{\nlone}{\ensuremath{N_L^{(1)}}\xspace}

\newcommand{\Xdat}[0]{\ensuremath{ {X^{\rm dat}}}\xspace}

\newcommand{\cppfid}[0]{\ensuremath{C^{\phi\phi, \rm fid}_L}\xspace}

\newcommand{\Cov}[0]{\ensuremath{\textrm{Cov}}\xspace} 
\newcommand{\av}[1]{\left \langle #1 \right \rangle}

\newcommand{\vn}[0]{\ensuremath{\boldsymbol{n}}}

\newcommand{\isdraft}[1]{#1}

\newcommand{\JC}[1]{{\isdraft{\color{purple} JC: #1}}}

\newcommand{\anorm}[0]{\ensuremath{\alpha}} 
\newcommand{\va}[0]{\ensuremath{\boldsymbol{\anorm}}\xspace} 

\newcommand{\hn}[0]{{\hat n}}
\newcommand{\Da}[0]{\ensuremath{\mathcal D_{\va}}} 

\newcommand{\D}[1]{\ensuremath{\mathcal D_{#1}}} 

\newcommand{\Beam}[0]{\ensuremath{\mathcal B}}

\newcommand{\Na}[0]{\ensuremath{N_{\va}}}
\newcommand{\Nai}[0]{\ensuremath{N^{-1}_{\va}}}

\newcommand{\Ma}[0]{\ensuremath{A_{\va}}} 

\newcommand{\Cova}{\ensuremath{\textrm{Cov}_{\va}}} 

\newcommand{\Rmfk}[0]{\ensuremath{\mathcal R^{\kappa\kappa, \rm MF}}\xspace}
\newcommand{\Rmfw}[0]{\ensuremath{\mathcal R^{\omega\omega, \rm MF}}\xspace}

\newcommand{\W}[0]{\ensuremath{\mathcal W}\xspace}
\newcommand{\Xunl}[0]{\ensuremath{X^{\text{unl}}}}

\newcommand{\XWF}[0]{\ensuremath{ {\hat{X}}_{\valpha}^{\text{WF}}}}

\graphicspath{{Figures/}}

\begin{document}

\title{Characterizing CMB noise anisotropies from CMB delensing}

\author{Louis Legrand}
\affiliation{Department of Applied Mathematics and Theoretical Physics, University of Cambridge, Wilberforce Road, Cambridge CB3 0WA, United Kingdom}
\affiliation{Kavli Institute for Cosmology, Cambridge, Madingley Road, Cambridge CB3 OHA, United Kingdom}

\author{Julien Carron}
\affiliation{Universit\'e de Gen\`eve, D\'epartement de Physique Th\'eorique et CAP, 24 Quai Ansermet, CH-1211 Gen\`eve 4, Switzerland}

\date{\today}

\begin{abstract}
Un-doing the effect of gravitational lensing on the Cosmic Microwave Background (`de-lensing') is essential in shaping constraints on weak signals limited by lensing effects on the CMB, for example on a background of primordial gravitational waves. Removing these anisotropies induced by large-scale structures from the CMB maps also generally helps our view of the primordial Universe by sharpening the acoustic peaks and the damping tail. However, practical implementations of delensing transfer parts of these anisotropies to the noise maps. This will induce a new large scale `mean-field' bias to any anisotropy estimator applied to the delensed CMB, and this bias directly traces large-scale structures.
This paper analytically quantifies this delensed noise mean-field and its impact on quadratic (QE) and likelihood-based lensing estimators.
We show that for Simons-Observatory-like surveys, this mean-field bias can reach 15\% in cross-correlation with large-scale structures if unaccounted for.
We further demonstrate that this delensed noise mean-field can be safely neglected in likelihood-based estimators without compromising the quality of lensing reconstruction or $B$-mode delensing, provided the resulting lensing map is properly renormalized.
    
\end{abstract}

\maketitle

\section{Introduction}

The light of the Cosmic Microwave Background (CMB) is deflected by the gravitational lensing due to the large scale structures~\cite{Lewis:2006fu}. 
Lensing thus deforms and blurs the image of the primordial CMB. This is of particular importance when trying to observe the primordial B modes, sourced by the gravitational waves created by inflation: constraints on the tensor to scalar ratio are limited by the variance from the secondary B modes created by lensing~\cite{Tristram:2021tvh,BICEP:2021xfz}.
To improve constraints on the tensor scalar ratio, delensing of the CMB maps is key~\cite{Carron:2017vfg,Belkner:2023duz,Namikawa:2021gyh,SimonsObservatory:2024gol}.
Delensing can also help in sharpening the peaks of the CMB temperature and polarization E modes spectra, which can improve the constraints on cosmological parameters~\cite{Green:2016cjr,Hotinli:2021umk}. 
Finally, delensing is at the core of the more efficient, likelihood-based CMB lensing methodology including Maximum A Posteriori (MAP) reconstruction~\cite{Hirata:2003ka,Carron:2017mqf,Legrand:2023jne,Legrand:2021qdu,Belkner:2023duz} or MUSE~\cite{Millea:2017fyd, Millea:2021had, SPT-3G:2024atg}.

By construction, delensing will reduce the statistical anisotropies in the observed CMB that were created by lensing. However, it will generate anisotropies in the CMB noise maps.
Indeed, even when assuming Gaussian and statistically isotropic noise maps, the delensing will deform this noise map in such a way that it will become statistically anisotropic.

These features in the noise maps can be characterized in CMB lensing terminology as a \textit{mean-field} -- a source of anisotropy potentially contaminating the signal of primary interest. Since this mean-field is introduced by delensing the noise maps, we will call it \textit{delensed noise mean-field}.

The scope of this article is to characterize this delensed noise mean-field. 
In Section~\ref{sec:delensednoisemf} we start by investigating qualitatively the delensed noise mean-field, and obtain a perturbative expression for it. Then in Section~\ref{sec:MAP} we investigate the impact of the delensed noise mean-field on the temperature only MAP estimator, and compare different ways of taking into account its contribution. We then check the impact of the delensed noise mean-field in delensing the B modes in Section~\ref{sec:delensing}, and we conclude in Section~\ref{sec:conclusion}. In Appendix \ref{app:MFresp} we detail the derivation of the delensed noise mean-field response.
 
\section{Delensed noise mean-field}
\label{sec:delensednoisemf}

\begin{figure}
    \centering
    \includegraphics[width=0.87\columnwidth]{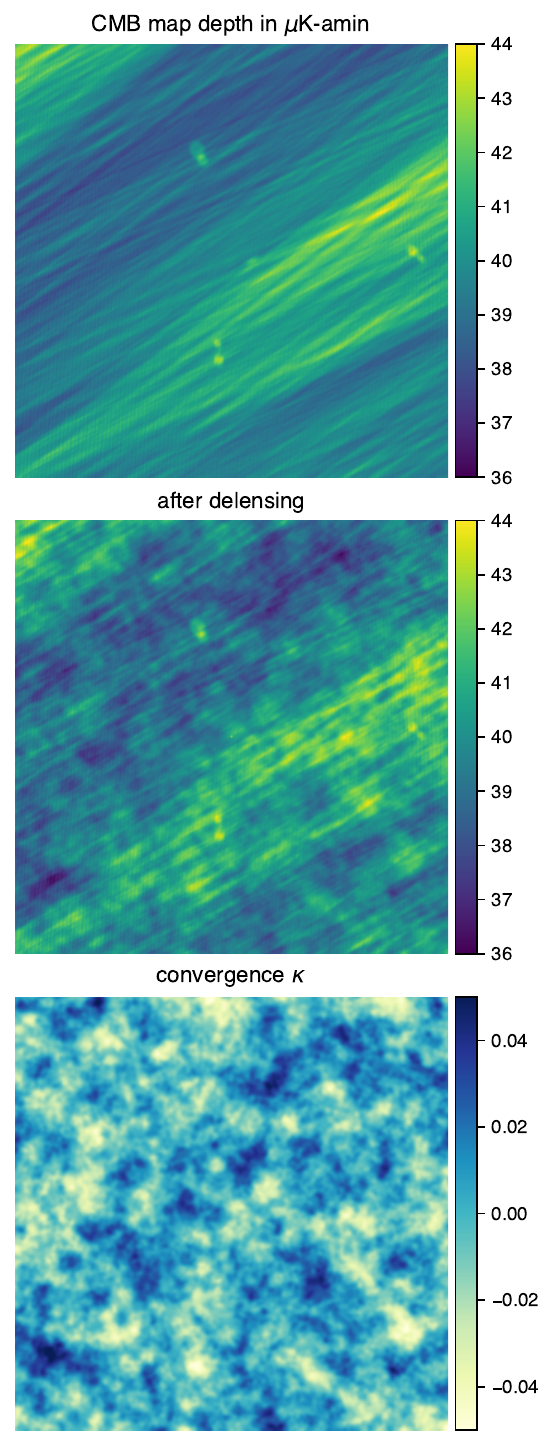}
    \caption{The top panel shows a randomly selected 7 degrees by 7 degrees patch of \emph{Planck} effective temperature depth map after SMICA component separation~\cite{Planck:2018yye, Carron:2022eyg}. Delensing using a Wiener-filtered lensing tracer with fidelity similar to that expected from the \emph{Planck} reconstruction will result in an effective depth given by the middle panel, where the original stripes caused by the scan strategy are modulated by the deflection field magnification. The resulting fluctuations trace the convergence map of the deflection field shown in the bottom panel.}
    \label{fig:delensnoise}
\end{figure}

\newcommand{\valpha}[0]{ {\boldsymbol{\alpha}}}
\newcommand{\Xdel}[0]{ {X^{\text{del}}}}
\newcommand{\Tdel}[0]{ {T^{\text{del}}}}

\newcommand{\vx}[0]{{\boldsymbol{x}}}
\renewcommand{\vl}[0]{{\boldsymbol{l}}}

In this section we provide a useful if slightly hand-wavy path towards the main results, neglecting all the complications of real-data. This is justified later, where we demonstrate rigorously that they also hold in more general contexts, including that of likelihood-based, `beyond-Quadratic-Estimator' lensing estimation in the presence of any sort of non-idealities.

A useful manner to think about delensing is simply remapping of the coordinate: let $\Xdat(\vx)= X^{\text{unl}}(\vx + \valpha(\vx)) + n(\vx)$ be the observed CMB Stokes fields plus noise, where \va is the lensing deflection field. We assume in this section for simplicity that the maps were successfully beam-deconvolved. Provided with any tracer of the lensing field $\hat \valpha$, one can simply attempt to remap the points 
\begin{equation}
	\Xdel(\vx) \equiv \Xdat(\vx - \hat \valpha(\vx))
\end{equation}(possibly after filtering the data, the tracer, or both). This is for example what the first delensing attempts by \cite{Larsen:2016wpa} and \cite{Carron:2017vfg} did \footnote{Note that this naive approach can lead to suboptimal results, see \cite{BaleatoLizancos:2020jby} for a detailed discussion on delensing templates.}. Reconstruction of the unlensed CMB based on a likelihood-model will perform the delensing in a more subtle manner, but we will see that the end result will be equivalent for our purposes -- we stick to this remapping picture in this section.  

Under this operation, the noise component is obviously also transformed, according to
\begin{equation}
	n(\vx) \rightarrow n(\vx - \hat \valpha(\vx)),
\end{equation}
and anisotropies will be induced in its covariance. 
Regions that were magnified by large scale lenses will be compressed to a smaller area, while preserving the surface brightness. Assuming the original instrumental noise was uncorrelated between pixels, this will reduce by some amount the effective local noise. The opposite happens in regions which were demagnified, where delensing will make the local noise go a bit up, by inflating the area by some amount.

We illustrate this in the Figure~\ref{fig:delensnoise}, where we show a patch of the effective temperature depth map from \emph{Planck}~\cite{Planck:2018yye, Carron:2022eyg}, and the resulting delensed noise depth map. We see that the noise variance map is modulated by the convergence field (which is related to the deflection field, $\kappa = -\nabla \cdot \valpha/2$.), creating anisotropies in the delensed noise map which are tracing the lensing convergence.

Consider then a standard quadratic estimator (QE) from temperature\cite{Okamoto:2003zw, Planck:2015mym}, applied to these delensed maps
\begin{align} \label{eq:qedel}
	\nonumber
	\bar x_{LM} &\equiv \frac 12 (-1)^M \sum_{\ell_1 m_1 \ell_2 m_2}\begin{pmatrix}
		L & \ell_1 & \ell_2  \\ -M &  m_1 & m_2 
	\end{pmatrix}\\ \times &  W^x_{L\ell_1 \ell_2}  \bar T^{\rm del}_{\ell_1 m_1}  \bar T^{\rm del}_{\ell_2 m_2}
\end{align}targeting some source of anisotropy $x$ from the signal part of these maps. For example residual lensing signal, or something else such as patchy reionization and cosmic birefringence \cite{Dvorkin:2008tf,Namikawa:2021gbr,BICEPKeck:2022kci,Darwish:2025fip}.
 Here the map $\bar T^{\rm del}$ is the inverse-variance filtered delensed $T^{\rm del}$ map. Ignoring any non-idealities other than lensing in this section, there are two main options. Either one may proceed in analogy to standard QE by ignoring the anisotropy in the covariance and writing
\begin{equation}\label{eq:qei}
	\bar T^{\rm del}_{\ell m } =  \frac{T^{\rm del}_{\ell m}} { C^{\rm del}_\ell + N_\ell}.
\end{equation}
where $C^{\rm del}_\ell$ is the delensed CMB signal spectrum and $N_{\ell}$ some choice of anisotropic noise spectrum. Or, one may take into account the fact that the CMB was delensed, and use instead `optimal-filtering',
\begin{equation}\label{eq:qef}
	\bar T^{\rm del} = \left(  C^{\rm del} + N_{\hat \valpha}\right)^{-1} T^{\rm del}
\end{equation}
where $N_{\hat \valpha}$ is now a dense matrix, accounting for the delensing-induced anisotropies in the noise.

In both cases, these estimators $\bar x$ will respond to the noise anisotropy induced by $\hat \valpha$ in the delensed maps, besides the sought-after signal. As usual, we can decompose the deflection field into a gradient ($\kappa = -\Delta \phi/2 $) and curl ($\omega = -\Delta \Omega/2$) component. We can then write schematically
\begin{align}\label{eq:RMF}
	\bar x_{LM} \ni \mathcal R^{x \kappa, \rm MF}_L \hat \kappa_{LM}  + \mathcal R^{x \omega, \rm MF}_L \hat \omega_{LM}  + \cdots
\end{align}
where the response function are defined as $\mathcal R^{x \kappa, \rm MF}_L = \partial \bar x_{LM}/\partial \hat \kappa_{LM} $ and $\mathcal R^{x \omega, \rm MF}_L = \partial \bar x_{LM}/\partial \hat \omega_{LM}$.
We have added the superscript $\text{MF}$ to these responses since in quadratic estimator language they are `mean-fields', i.e. a bias in the estimated map $\bar x$, rather than sought-after anisotropic signals in the CMB maps.

The responses for the filtering choices \eqref{eq:qei} and \eqref{eq:qef} will be different. However, to leading order in $\hat \valpha$, one has simply
\begin{align}
	& \mathcal R^{x \kappa, \rm MF} \text{        (isotropic filtering, \eqref{eq:qei})} \nonumber\\
	 = 	-&\mathcal R^{x \kappa, \rm MF} \text{      (anisotropic filtering, \eqref{eq:qef})}
\end{align}
and similarly for $\omega$.
This is shown in details in appendix~\ref{app:MFresp}, but it is not too difficult to see why that can be: regions where the local noise variance is made higher (lower) by delensing will be accordingly down-weighted (up-weighted) by the optimal filtering~\eqref{eq:qef}, resulting eventually in an anisotropy of the opposite sign in the filtered map. From now on, the sign of our mean-field responses refer to optimal filtering~\eqref{eq:qef}. The reason is that this corresponds to the case of beyond-QE, likelihood-based lensing reconstruction, which always filters maps using all information available.
\begin{figure}
    \centering
    \includegraphics[width=0.99\columnwidth]{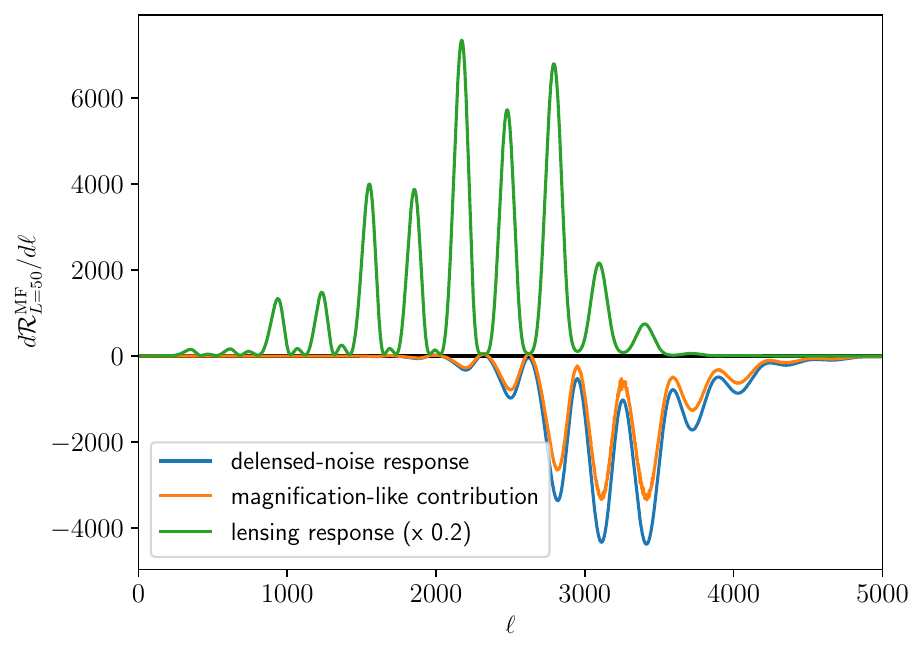}
    \caption{The blue line shows the contribution per multipole to the large scale (here $L=50$) delensed noise mean-field response to the lensing convergence map, for a SO-like configuration, with a Gaussian beam of $3$-arcmin and white noise. The squeezed approximation in \eqref{eq:squeezk} is a perfect match, with the orange curve highlighting the contribution from converging rather than shearing effects on the delensed noise map. The green line shows for comparison the standard lensing signal contributions to the $TT$ lensing quadratic estimator (reduced by a factor 5). The delensed noise mean-field is sourced from significantly smaller scales, where noise is relevant but not completely dominating either. This is for temperature reconstruction.}
    \label{fig:dmfdl}
\end{figure}

The response of a standard QE defined as in \eqref{eq:qedel}, but applied on temperature maps without delensing, is given by \cite{Okamoto:2003zw,Planck:2015mym}
\begin{equation}
	\label{eq:standardresponse}
	\mathcal{R}^{x \kappa}_L = \frac{1}{2(2L+1)} \sum_{\ell_1 \ell_2} f^{TT, \kappa}_{L \ell_1 \ell_2} W^x_{L \ell_1 \ell_2} F^T_{\ell_1} F^T_{\ell_2} \, .
\end{equation}
where $f^{TT, \kappa}$ is the response of the pair of fields $TT$ to the lensing convergence, as given in Table 1 of \cite{Okamoto:2003zw}, and $F^T_{\ell}$ is the inverse variance filtering approximated as diagonal in $\ell$. 

To obtain the delensed noise mean-field response $\mathcal R^{x \kappa, \rm MF}_L$, we make two modifications to this fiducial response (see Appendix~\ref{app:MFresp} for a detailed derivation).
First, we replace the filters by $F_\ell= 1/(C^{\rm del}_\ell + N_\ell)$, to match the delensing filtering as in Eq.~\eqref{eq:qei} and Eq.~\eqref{eq:qef}.
Second, since this mean-field is sourced by the noise, we replace the lensed CMB spectra (or the lensed gradient spectra, see \cite{Lewis:2011fk,Maniyar:2021msb}) entering the response $f$ by the CMB noise spectra $N_\ell$ .
The QE weights $W$ depend on the choice of QE made. In the Section \ref{sec:MAP}, with the MAP estimator, we will use delensed spectra $C^{\rm del}_\ell$ in the weights $W$ instead of the lensed (or lensed gradient) spectra.

Since we expect the delensed noise mean field to follow the deflection field, whose power spectrum peaks at large scales, $L\sim 30$, we now discuss the squeezed limit of the response. Let $\Rmfk_L$ and $\Rmfw_L$ be the linear response of the convergence $\kappa$ estimator mean-field to $\kappa$, and of the field rotation $\omega$ estimator mean-field to $\omega$, respectively (as always, from parity arguments cross-terms vanish). Since the only difference to a standard QE response is that the anisotropy source enters the noise, we can use as starting point the squeezed limit of the standard estimator and make the corresponding modifications. The standard estimator response at low $L$ is~\cite{Bucher:2010iv, Carron:2024mki}

\begin{align}
	\label{eq:squeez}
				&\mathcal R^{\kappa \kappa}_L \sim \frac 12 \sum_\ell \left(\frac{2\ell + 1}{4\pi} \right) \W^2_\ell\: \times \\ & \left[\left(\frac{d  \ln \ell^2 C_\ell}{d\ln \ell} \right)^2 
		 + \frac 12  \Gamma_L^2 \left(\frac{d  \ln C_\ell}{d\ln \ell}\right)^2\right] \nonumber
\end{align}
where \W is the isotropic Wiener-filter
\begin{equation}
	\W_\ell \equiv \frac{C_\ell}{C_\ell + N_\ell}.
\end{equation}
and
\begin{equation}
	\Gamma_L^2 \equiv \frac{(L+2)(L - 1)}{L(L +1)}
\end{equation}
is the shear factor.

The first term in brackets is the information from local magnification (vanishing for scale-invariant spectra), and the second from shear (vanishing for white spectra). The $L$-dependent shear prefactor is only relevant at the very lowest multipoles.
The mean-field response is found by realizing that in the general case, one factor of $C_\ell$ in the numerators is the fiducial spectrum used as QE weight, and the second is the spectrum affected by the anisotropy. Hence, the correct result is then found by replacing one of the two $C_\ell$ by $N_\ell$.
Doing so, one obtains
\begin{align}
	\label{eq:squeezk}
				&\Rmfk_L \sim \frac 12 \sum_\ell \left(\frac{2\ell + 1}{4\pi} \right) \W_\ell (1 - \W_\ell)\: \times \\ & \left[\frac{d  \ln \ell^2 C^{\rm del}_\ell}{d\ln \ell}\frac{d \ln \ell^2 N_\ell}{d\ln \ell}  
		 + \frac 12 \Gamma_L^2\frac{d  \ln C^{\rm del}_\ell}{d\ln \ell}\frac{d \ln N_\ell}{d\ln \ell} \right] \nonumber
\end{align}
with
\begin{equation}
		\W_\ell \equiv \frac{C^{\rm del}_\ell}{C^{\rm del}_\ell + N_\ell}.
\end{equation}
The factor $\W_\ell (1 - \W_\ell)$ in Eq.~\eqref{eq:squeezk} selects now the CMB modes where noise is relevant, but not too noisy so as to be useless. The factor $(1 - \W_\ell)$ appears by replacing the $C_\ell$ by $N_\ell$ in the numerator of $\W_\ell$, and corresponds a Wiener filter of the noise.

Further, from the same arguments, the lensing rotation response from large lenses is identical to the shear-part of the lensing convergence response from large lenses:
\begin{equation}
	\label{eq:squeezw}
	\begin{split}
		\Rmfw_L & \sim \frac 12 \sum_\ell \frac{2\ell + 1}{4\pi} \W_\ell (1 - \W_\ell) \: \times \\ 
		& \left[\frac 12 \Gamma_L^2 \frac{d  \ln C^{\rm del}_\ell}{d\ln \ell}\frac{d \ln N_\ell}{d\ln \ell} \right]
	\end{split}
\end{equation}
The reason for this is as follows: in contrast to the (de-)magnifying effect of a constant convergence, a constant rotation on a patch of the CMB temperature has no locally observable effect at all. Hence, it is the contribution of $\omega$ to the shear $B$-mode responsible for Eq.~\eqref{eq:squeezw}, which is the same than that of $\kappa$ to the shear $E$-mode contribution to Eq.~\eqref{eq:squeezk}.

For high-resolution experiments, a white noise model is often accurate over a wide range of scales. Hence, the limit of perfectly white noise is both simple and of some relevance. In this case only the convergence piece contributes, and can be simplified further\footnote{To get \eqref{eq:sqzwhite}, we used  $(1- \W) d \ln C = d \ln \W$, which holds for white noise, and integrated by part. This limit does match the estimate obtained independently combining B.9 and B.12 of \cite{Belkner:2023duz}, and improves slightly the one given in the main text of that reference.} to the following result,
with $\ell_{\text{max}}$ the maximum multipole used in the CMB temperature maps.
\begin{align}\label{eq:sqzwhite}
	\Rmfk_L &\sim  - 2\sum_\ell  \frac{2 \ell + 1}{4\pi}(\W_\ell^2 - \W_{\ell=\ell_{\text{max}}})  \\
	\Rmfw_L &= 0 \quad \text{ (low $L$'s, white noise)}
\end{align}
This shows that as for the standard lensing response, the dependence on $\ell_{\text{max}}$ can be very strong. If the range of CMB multipoles used is large enough and enters the noise dominated regime, the response will saturate to $-2\sum \W_\ell^2 (2\ell + 1) /4\pi$. On the other hand, it is always zero in the regime where all the modes considered are noise-free ($\W_\ell \sim 1$), which is quite relevant for deep experiment like CMB-S4 for example.

Realistic noise is of course not perfectly white, but grows sharply on scales below the resolution of the experiment. This can make the shear-type contribution very substantial. For calculations we use a standard Gaussian model for which
\begin{equation}
\ln N_\ell = \left( \frac{\sigma_{\rm FWHM}}{8 \ln 2 }\right)^2 \ell (\ell + 1) + \text{const.}
\end{equation}
where $\sigma_{\rm FWHM}$ is the beam full width at half maximum. The contribution per mode to the delensed noise mean-field response is shown in Figure \ref{fig:dmfdl} for a SO-like configuration.

\begin{figure}
    \centering
    \includegraphics[width=0.99\columnwidth]{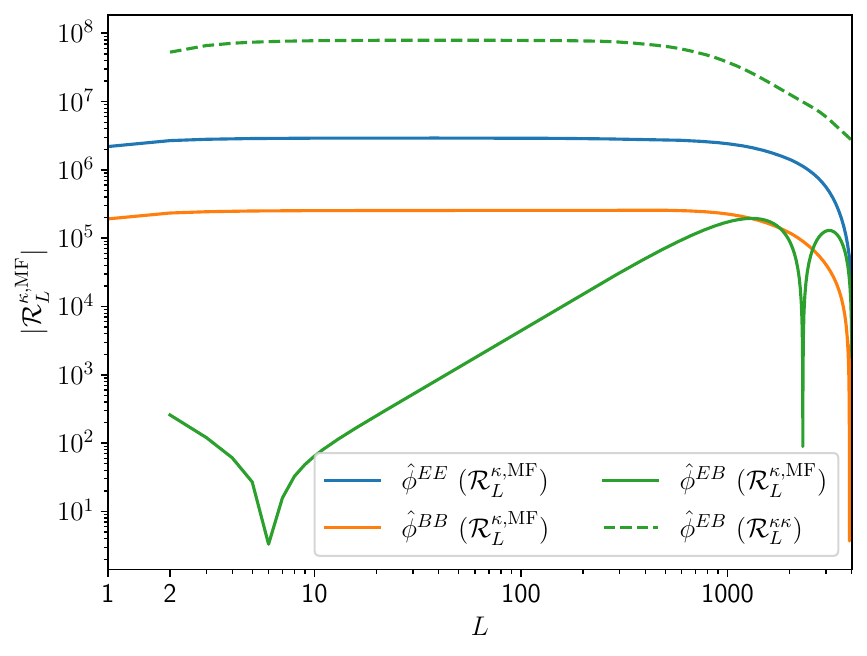}
    \caption{Delensed-noise mean-field response for the polarized lensing quadratic estimator, for a configuration approaching crudely the planned CMB-S4 wide survey, with equal power in the $E$ and $B$ noise spectra. All curves were computed using the lensed CMB spectra. In contrast to standard lensing, the $EB$ mean-field response is essentially zero (green), owing to the symmetry in unlensed $E$ and $B$ noise power, and $EE$ (blue) dominate the response, followed by $BB$ in orange. Dashed green shows the standard lensing response in the same configuration for comparison.}\label{fig:polresp}
\end{figure}

Let us now discuss briefly the case of polarization. For lensing on intermediate scales, e.g. $\ell_{\rm max} = 3000$, and low instrumental noise, it is well known that the $EB$ estimator carries most of the signal. Large lenses creates locally a $EB$ spectrum proportional to $C_\ell^{EE} - C_\ell^{BB}$~\cite[see e.g.][]{Maniyar:2021msb}. Since the primordial $C_\ell^{BB}$ spectrum is very weak, this makes this signal very significant, provided a sufficient number of modes is resolved. On the other hand, the $EE$ estimator behaves similarly to temperature (the magnification and shearing effects discussed above), while the $BB$ lensing estimator is near irrelevant.

Lensing acting on the noise maps is different, since the strong $E$-$B$ asymmetry of the primordial CMB signal is lost. Since $N^{EE}$ is comparable to  $N^{BB}$, there is no creation of small-scale $EB$ power, and the $EB$ mean-field response will vanish on large scales. Both $EE$ and $BB$ will behave now like the temperature case \eqref{eq:squeezk}. The relative weight of the mean-field response compared to the $EB$ response is much weaker, so that the overall effect in low-noise polarization reconstruction is always very small. Likewise, this should only have a very small impact on the $TE$ and $TB$ estimators, which probes the unlensed $TE$ spectrum. Figure~\ref{fig:polresp} shows the polarization mean-field responses for a CMB-S4 like configuration.

\section{Impact on optimal-lensing reconstruction}
\label{sec:MAP}

We now investigate the role and impact of the delensed noise mean-field on likelihood-based lensing reconstruction. First proposed by~\cite{Hirata:2003ka}, usage of the CMB-lensing likelihood allows better (in principle, optimal) reconstruction of the CMB lensing deflections. 

We start in Section~\ref{sec:margMAP} discussing how this mean-field enters the original approach of~\cite{Hirata:2003ka}, which was revisited more recently by~\cite{Carron:2017mqf}. See \cite{Belkner:2023duz} for a detailed description of the resulting algorithm in spherical geometry.

Ref.~\cite{Millea:2017fyd} introduced the alternative approach of using the joint primordial CMB and CMB lensing likelihood, whereas \cite{Hirata:2003ka} uses the one marginalized over the primordial CMB fields. This approach is also used by the MUSE framework~\cite{Millea:2021had}, and was employed by the latest SPT-3G cosmological analysis~\cite{SPT-3G:2024atg}.
The marginal and joint approaches are related in a precise manner: this second approach produces the same CMB and lensing maps than the first, were the first method to neglect all sources of mean-fields, as discussed in Ref.~\cite{Carron:2025wqb}. In particular, the delensed noise mean-field. Hence our results are also relevant for the MUSE approach, since the maps produced by this method will contain this mean-field\JC{\footnote{In the MUSE framework, a prior on the mean convergence value is introduced in order to mitigate the impact of the mean-field~\cite{Millea:2021had}.}}.

We then test the quantitative impact of this mean-field on temperature-based reconstructions in Sections \ref{sec:tests} to \ref{sec:normalization}. Finally in Section \ref{sec:delensing}, we investigate the impact of the delensed noise mean-field on the delensing of the $B$ modes.
In this Section we work with the FFP10 cosmology\footnote{\url{https://github.com/carronj/plancklens/blob/master/plancklens/data/cls/FFP10_wdipole_params.ini}}.
Note that we work at fixed cosmology, i.e. we do not evaluate the impact of using a non-fiducial cosmology in the simulations. In \cite{Legrand:2021qdu,Legrand:2023jne}, we found that the the impact of the cosmology used for the lensing reconstruction was negligible for the iterative lensing normalization, and the lensing biases (\nlzero and \nlone) can be mitigated by using a realization-dependent debiaser. In a end-to-end cosmological analysis like in \cite{Legrand:2021qdu}, we can then include a first order correction on the response and on the biases with respect to the sampled lensing spectrum and CMB spectra, following \cite{Planck:2018lbu}. Additionally, in \cite{Belkner:2023duz}, we found that the fiducial cosmology does not bias the recovered tensor to scalar ratio $r$ from $B$-mode delensing. Thus we do not expect the delensed noise mean-field to be impacted by a non-fiducial cosmology used in the simulations.

\subsection{MAP estimator and delensed noise mean-field}\label{sec:margMAP}

\begin{figure*}
	\centering
	\includegraphics[width=\textwidth]{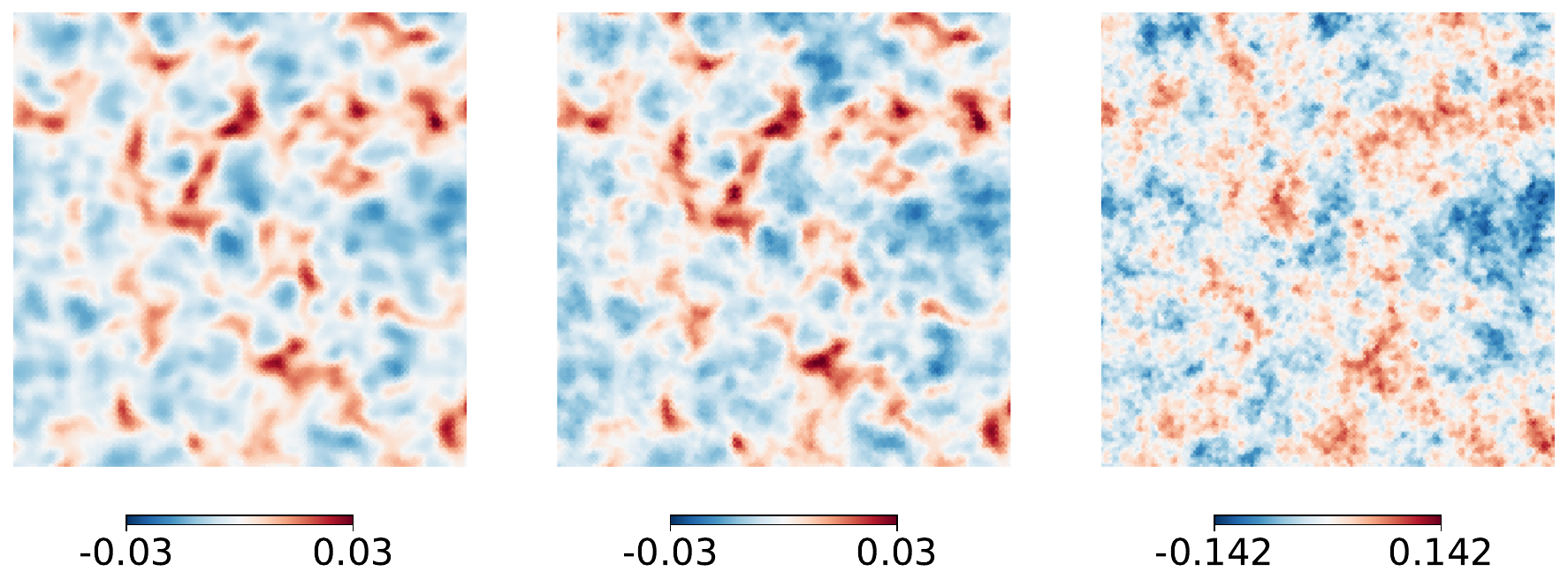}
	\caption{Estimation of the delensed noise mean-field $\kappa^{\rm MF}$ in real space, at the first iteration of the MAP estimator from temperature. Simulations are with CMB-S4 like noise level. The left panel shows the predicted delensed noise mean-field at first order in the deflection field. The central panel shows the delensed noise mean-field estimated from a set of 20 simulations at fixed deflection field. The right panel shows the negative of the Wiener filtered true convergence field of the simulation. Patches are 200x200 pixels, with 1.5 arcmin per pixel, cutouts from full-sky simulations.}
	\label{fig:mf_maps}
\end{figure*}

The key ingredient of likelihood-based reconstruction is a good model for the CMB covariance matrix conditioned on the deflections.
Let the observed CMB Stokes fields be \Xdat,  and \Cova~the covariance. The simplest covariance matrix models will be of the form
\begin{equation}\label{eq:cov}
	\begin{split}
		\Cov_{\va} &\equiv
		\Beam \Da \mathcal{Y} \:C^{\rm unl} \mathcal{Y}^\dagger \Da^\dagger \Beam^\dagger + N \, .
	\end{split}
\end{equation} 
where $\Da$ is the lensing remapping operation ($X^{\rm len} = \Da X^{\rm unl}$, where $\Xunl$ is the unlensed CMB), $\Beam$ is the beam of the experiment, $\mathcal{Y}$ is compact notation for the spherical harmonic transforms that maps spherical harmonics coefficients of the unlensed CMB to the real space maps, $C^{\rm unl}$ is the covariance matrix of the unlensed CMB, which is diagonal in harmonic space, and $N$ the noise covariance matrix.

The deflection field posterior is then
\begin{equation}
    \label{eq:logpost}
	\begin{split}
     -2   \ln \mathcal{P}^{\rm M}(\va | \Xdat)& = \Xdat^\dag \Cov_{\va}^{-1}\Xdat + \ln \det \Cov_{\va} \\ 
     &+ \kappa \cdot \:C^{\kappa\kappa, -1} \kappa.   
	\end{split}
\end{equation}
The term
\begin{equation}
	\label{eq:prior}	
	\kappa \cdot \:C^{\kappa\kappa, -1} \kappa \equiv \sum_{LM} \frac{|\kappa_{LM}|^2}{C_L^{\kappa\kappa}}
\end{equation}
is a Gaussian prior on the lensing map, which plays no role in this discussion.
The likelihood gradient at non-zero deflection field $\valpha$ contains the information on the residual lensing signal. Explicitly, one can define~\cite{Carron:2017mqf}
\begin{equation}
    \label{eq:gradtot}
    \begin{split}
 &\frac{\delta }{\delta \alpha_a(\hat n)}\left[ \ln \mathcal{P}^{\rm M} ( \va | \Xdat )\right] 
        \\& \equiv \:_{a}{g_{\va}^{\rm QD}}(\hat n) - \:_{a}{g_{\va}^{\rm MF}}(\hat n) + \:_{a}{g_{\va}^{\rm Pri}}(\vn)  \, , 
    \end{split}
\end{equation}
The terms $g_{\va}^{\rm QD}$ and $g_{\va}^{\rm MF}$ are the piece quadratic in the data and the mean-field respectively, and the subscript $a$ denotes one of the two coordinates directions on the sphere. The quadratic gradient is the piece that tries and capture residual lensing signal, beyond that represented by $\valpha$, the current estimate of the deflection field. For example, $g_{\valpha = 0}^{\rm QD}$ is quite literally the standard quadratic estimator~\cite{Okamoto:2003zw}, prior to normalization, in the generalized minimum variance (GMV) form discussed by \cite{Maniyar:2021msb}. For non-zero $\valpha$, the form of $g_{\valpha}^{\rm QD}$ is only slightly different, with the differences accounting for
\begin{enumerate}
	\item first constructing the delensed CMB (`$\XWF$'), under the assumption that $\valpha$ and the other ingredients of the likelihood model are the truth
	\item building then a quadratic estimator from these partially delensed maps, that will capture residual anisotropies.
\end{enumerate}
More precisely, the quadratic piece may be written in the following form (see~\cite{Carron:2025wqb}),
\begin{equation}\label{eq:gdel}
		g_{\va}^{\rm QD}(\hn) = |\Ma|(\hn)\left[ \Da g_{\va}^{\rm QE, \rm del}\right](\hn),
\end{equation}
which is a useful representation in order to understand the mean-field: in this equation, $g_{\va}^{\rm QE, del}$ can be unambiguously identified to a standard quadratic estimator on delensed maps, as defined in~\cite{Carron:2025wqb}. This quadratic estimator is of the `optimal filtering' type, \eqref{eq:qef}, which includes delensed noise anisotropies in the filters. The prefactor $|A_{\valpha}| = |d^2\hn'/ d^2\hn| \sim 1 - 2\kappa$ is the magnification matrix determinant of the sphere remapping $\hn \rightarrow \hn'$ induced by $\valpha$. 
The logic behind the factors $|A_{\valpha}|(\hn) \Da$ in Eq.~\eqref{eq:gdel} is this: $g_{\va}^{\rm QE, \rm del}$ captures residual lensing on the delensed maps, hence probes residual deflections at the \emph{unlensed} positions, not the observed ones. These factors, that combine the remapping operation together with the distortions of the coordinates from unlensed to observed space, are precisely the ones obtained by the chain rule that connect the likelihood gradients with respect to the deflection field at the unlensed and lensed positions.

The second term in Eq.~\eqref{eq:gradtot} corresponds to a mean-field, it is the gradient of the log-determinant of the covariance. It is independent of the data, by definition. 
By noting that the first variation of the log-likelihood should be zero on average, we can write the mean-field gradient as 
\begin{equation}
	g_{\va}^{\rm MF} = \left< g_{\va}^{\rm QD} \right>_{\text{fixed }\va} 
\end{equation}
where the average occurs over realizations of the data model \eqref{eq:cov} conditioned on $\va$.
According to Eq.~\eqref{eq:gdel}, this is proportional to a mean-field of a standard QE. It follows that 

\begin{enumerate}
	\item in the absence of anisotropies other than lensing, since the QE mean-field $\av{g_{\va=0}^{\rm{QE, del}}}$ vanishes, we can simply drop the prefactor when working to linear order in $\va$, with the result
\begin{equation}
		g_{\va}^{\rm MF} \sim \av{g_{\va}^{\rm QE, \rm del}}_{\text{fixed }\va}.
\end{equation} This shows that the mean-field linear (perturbative) response $\mathcal R_L^{\rm MF}$ to $\valpha$ will be that of a QE with anisotropic filtering, defined just as we did in the previous section in Eq.~\eqref{eq:RMF}.
\item In practice, the zeroth-order QE mean-field will never be zero. This part then just gets remapped, and modulated by the magnification.
\end{enumerate}

We note that the perturbative prediction of the mean-field from Eq.~\eqref{eq:RMF} is equivalent to modifying the prior on the lensing field in Eq.~\eqref{eq:prior}. Indeed, the perturbative mean-field can be absorbed in the prior which becomes 
\begin{equation}
	\kappa \cdot \: C^{\kappa\kappa, -1} \kappa  \rightarrow \kappa \cdot \:(C^{\kappa\kappa, -1} - \mathcal{R}^{\kappa\kappa, \rm MF}) \kappa \, .
\end{equation}
As we will discuss in Section~\ref{sec:normalization}, the predicted normalization of the MAP estimator is a Wiener filter driven by the prior. Thus neglecting the delensing mean-field will change the normalization of the MAP estimator.

We show on the Figure~\ref{fig:mf_maps} an estimate of the delensed mean-field at the first iteration of the MAP estimator. Thus the deflection field $\va$ on this figure is obtained from the quadratic estimator. 
We show on the left panel the mean-field $g_{\va}^{\rm MF}$ from the perturbative expression of Eq.~\eqref{eq:RMF}, and on the central panel the mean-field estimated from averaging over 20 simulations at fixed deflection field \va. 
We see that the perturbative mean-field is in good visual agreement with the one estimated from the simulations.
The right panel shows that the delensed noise mean-field is indeed correlated with the true lensing field, as expected.

\subsection{Mean-field tests}\label{sec:tests}

\begin{table}
    \begin{center}
		\begin{tabular}{ c  c  c  c }
			Experiment &  Planck & Simons Obs.  & CMB-S4  \\
			\hline
			\hline
			Noise $[\mu\rm K . \rm arcmin]$ & 35 & 3 & 1 \\
			Beam FWHM $[\rm arcmin]$ & 5 & 3 & 1 \\
			$\ell_{\rm min}$ & 100 & 40 & 30 \\
			$\ell_{\rm max}$ & 2048 & 4000 & 4000 \\
		   \hline   
		  \end{tabular}
    \end{center}
    \caption{Experimental configuration of our simulations.}
    \label{tab:surveys}
\end{table}

When reconstructing the lensing field with the MAP estimator, the mean-field gradient term should in principle be estimated at each iteration of the algorithm. 
However, performing Monte-Carlo estimation of this mean-field at each iteration is computationally expensive and induces residual Monte-Carlo noise in the gradient. Hence there is value in exploring other possibilities to treat this term.
We consider different schemes to see the impact of the MF gradient term in the iterations.

\begin{description}
	\item[Neglecting the MF] We neglect the MF and perform the lensing iterations assuming $g^{\rm MF}_{\va}=0$. In practice, this corresponds to truncating the likelihood gradient in Eq.~\eqref{eq:gradtot} during the MAP search. As such the lensing potential at convergence might not correspond to the true maximum of the posterior. However it has the advantage of being computationally cheap. This also matches the implementation of the CMB-joint MAP estimator used in MUSE, which does not include the MF term.
	
	\item[Simulated MF] We estimate the MF using the relation $g^{\rm MF}_{\va}=\left<g^{\rm QD}_{\va} \right>$. In practice, for each iteration in the MAP search, we estimate the average of the quadratic gradient term $g^{\rm QD}_{\va}$ over a set of 20 simulations, lensed by the estimated deflection field $\va^{\rm MAP}$. 
	We decrease the variance of this MF estimate by using the tricks of the appendix B2 of \cite{Carron:2017mqf}. Specifically, we modify the quadratic gradient term to operate on maps with unit variance instead of the CMB power spectrum variance, while preserving the expected value. Additionally, we further reduce variance by subtracting an MF estimate derived from the same set of unlensed CMB simulations.
	To remove any potential bias, at each iteration we use a different set of simulations to estimate the mean-field. 
	
	\item[Perturbative MF] We estimate the delensed noise mean-field using the perturbative approach described in Section~\ref{sec:delensednoisemf}.  The MF estimate is thus given by $g^{\rm MF}_{\va, \LM}= \mathcal R^{\kappa\kappa, \rm MF}_L \hat \kappa^{\rm MAP}_\LM$, where $\mathcal R^{\kappa\kappa, \rm MF}$ is obtained from Eq.~\ref{eq:standardresponse} using the noise spectra in the CMB response and delensed CMB spectra in the QE weights.
	
\end{description}

We consider three different CMB configurations, reproducing the Planck, Simons Observatory and CMB-S4 white noise. The corresponding beam, noise levels and CMB scale cuts are described in Table~\ref{tab:surveys}. 
We generate ideal full-sky simulations, without any sources of anisotropies other than the lensing field itself. In practice we generate a Gaussian unlensed CMB temperature map, lens it with a Gaussian lensing field, and add a realization of the noise. 
We reconstruct the lensing field on the full-sky with the MAP estimator as described in details in \cite{Belkner:2023duz}. 
We rely only on the temperature map to perform the lensing reconstructions.

\subsection{Impact on lensing reconstruction}

\begin{figure}
	\centering
	\includegraphics[width=0.95\columnwidth]{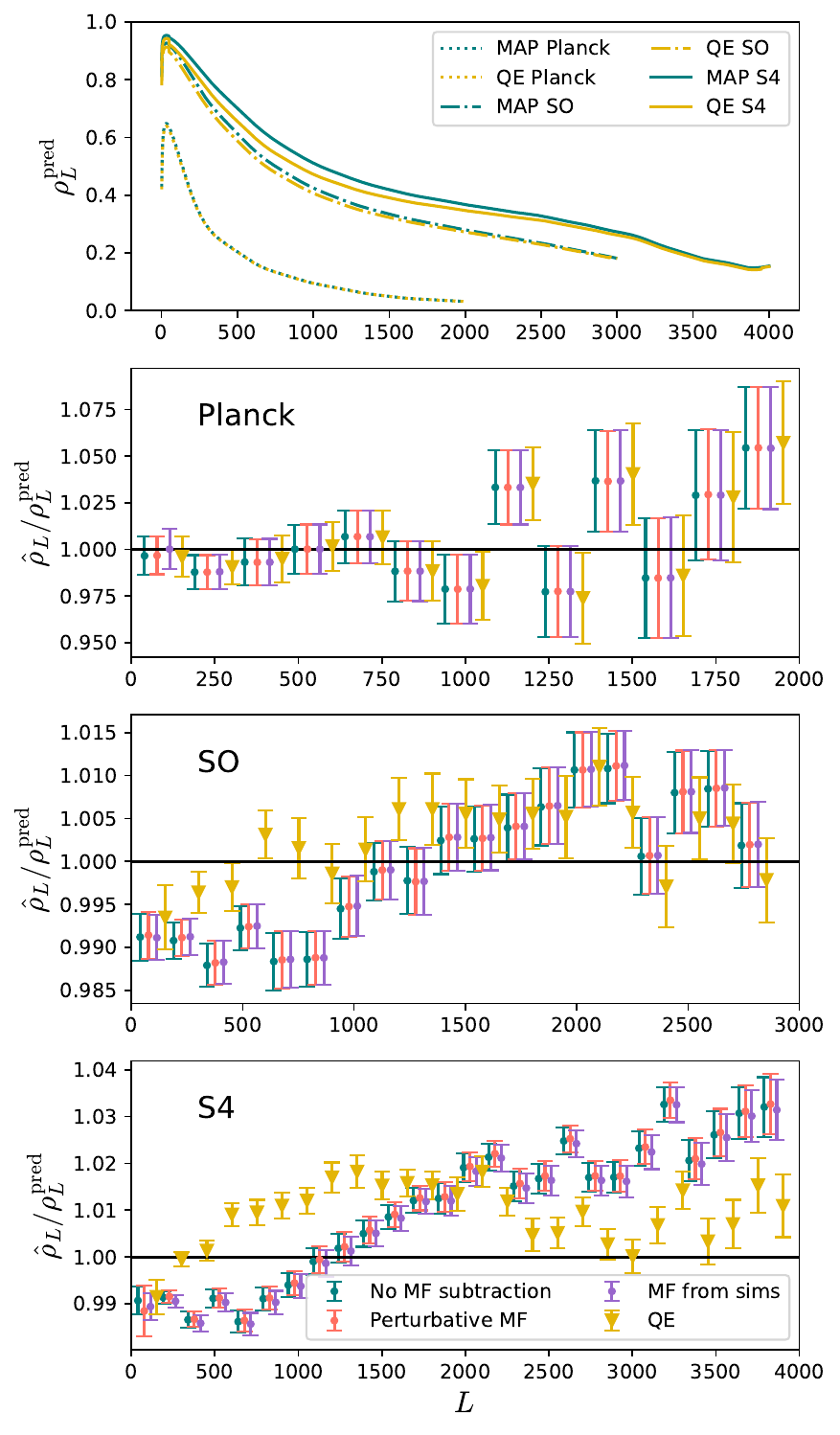}
	\caption{Upper panel: Predicted correlation coefficients for the temperature only MAP (teal) and QE (golden) reconstruction for Planck, SO and CMB-S4 like noise levels, respectively in dotted, dash-doted and plain lines.
	Lower panels: Correlation coefficient after 10 iterations of our lensing reconstruction divided by the fiducial correlation coefficient, for Planck, SO and CMB-S4 noise levels from top to bottom. 
	Note that the range of $L$ value is not the same in each panel. 
	In teal we show the case without mean-field subtraction, in pink it is using the perturbative mean-field prediction and in purple it is with the MF estimated from simulations. The QE case is shown in golden for comparison. 
	We see that for all recipes we find very similar correlation coefficients: neglecting the delensed noise mean-field has no impact on the quality of the lensing potential reconstruction. We also see that the predicted correlation coefficients are accurate at $\sim 2 \%$.}
	\label{fig:corr_ratio}
\end{figure}

For each experimental setting, we estimate the correlation coefficient between the reconstructed MAP lensing field $\hat \phi^{\rm it}$ and the simulation input lensing field $\phi^{\rm in}$
\begin{equation}
	\rho_L^{\rm MAP} = \frac{C_L^{\hat \phi^{\rm it}, \phi^{\rm in}}}{\sqrt{C_L^{\hat \phi^{\rm it}, \hat \phi^{\rm it}}C_L^{\phi^{\rm in}, \phi^{\rm in}}}} \, .
\end{equation}
The predicted correlation coefficient is given by
\begin{equation}
	(\rho^{\rm fid}_L)^2 = \frac{\cppfid}{ \cppfid + \nlzero + \nlone} \, ,
\end{equation}
where \nlzero and \nlone are the iterative lensing reconstruction biases, estimated from the fiducial procedure of \cite{Legrand:2021qdu,Legrand:2023jne}.

Figure~\ref{fig:corr_ratio} compares the lensing correlation coefficient for our different MF scenarios, after 10 iterations of the MAP estimator.
The predicted correlation coefficient is shown in the upper panel, for the MAP (teal) and QE (golden) reconstructions. The lower panels show the correlation coefficient after 10 iterations of our simulations, normalized by the fiducial correlation coefficients, for Planck, SO and CMB-S4 noise levels from top to bottom.
We bin the correlation coefficient in $L$-bins of size $\Delta L = 150$, showed with offset for clarity. Error bars are the standard error on the mean of the correlation coefficient in each bin.

The agreement with the predicted correlation coefficient is at the 2\% level. The prediction of the \nlzero and \nlone iterative biases are not perfect, as discussed in \cite{Legrand:2021qdu,Legrand:2023jne}, since they are obtained from a heuristic approach, and this is probably the main source of discrepancy. The bias in the QE correlation coefficient for CMB-S4 is also at the 2\% level, this might be due to the fact that we work on the full-sky, and the \nlone prediction is obtained in the flat-sky approximation.

We see that using a different recipe for the delensed noise mean-field, either perturbative or from the simulations, or completely neglecting it, has no impact on the correlation coefficient at all scales. 

We note that for the lowest multipole bin, the perturbative MF has a larger error bar than the simulated MF. This can be due to the fact that the perturbative approximation breaks down, as the approximation is only valid at first order in the deflection field. However, this does not impact the correlation coefficient itself.

On top of these tests, we also looked at neglecting the delensed noise mean-field in all iterations but the last one, or to use a simulated MF with a filtering of the small scales (which contain Monte-Carlo noise). We found that these methods do not improve over the simple case of neglecting the mean-field in all iterations.

\subsection{Power spectrum normalization}
\label{sec:normalization}

\begin{figure}
    \centering
    \includegraphics[width=\columnwidth]{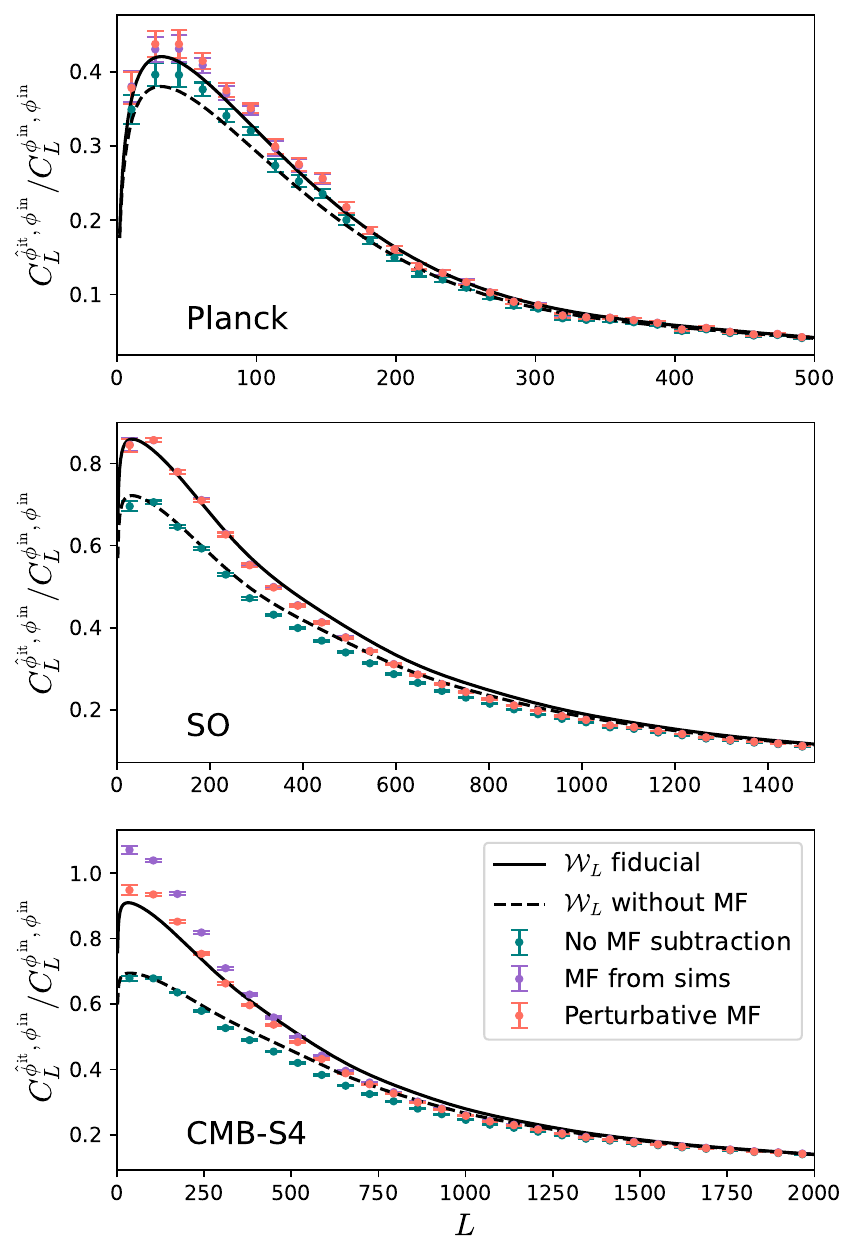}
    \caption{Isotropic normalization of the MAP lensing potential, for the Planck, SO and S4 noise levels from top to bottom. In teal we show the case without mean-field subtraction, in purple it is with the mean-field estimated from simulations, and in pink it is when using the perturbative mean-field prediction. The black plain lines is the fiducial MAP response and the dashed black line is the fiducial MAP response when we neglect the delensed noise mean-field in the likelihood.}
    \label{fig:mf_so_wf}
\end{figure}

We now consider the impact of the delensed noise mean-field on the normalization of the MAP lensing potential. 

We recall that our best prediction for the MAP lensing power spectrum normalization is given by a Wiener filter
\begin{equation}
	\label{eq:WF}
	\mathcal{W}_L = \frac{C^{\phi\phi, \rm fid}_L}{C^{\phi\phi, \rm fid}_L + N_L^{(0)}} \, ,
\end{equation}
where $C^{\phi\phi, \rm fid}_L$ is the fiducial lensing power spectrum and $N_L^0$ is the iterative $N_L^0$ bias. However we showed in \cite{Legrand:2021qdu,Legrand:2023jne} that this prediction is not accurate, and has to be corrected from a set of simulations. In practice for the polarization estimator we found that the prediction is off by around 3\%. Crucially, this correction is independent of the  CMB and noise power spectra contained in the map: it only depends on the fiducial ingredients used for the MAP reconstruction.

We show in Figure~\ref{fig:mf_so_wf} the effective normalization for the MAP lensing field with the temperature only estimator. This effective normalization is estimated from the cross correlation between the MAP lensing field $\hat\phi^{\rm it}$ after 10 iterations, with the true lensing field of the simulation $\phi^{\rm in}$
\begin{equation}
	\mathcal{\hat W}_L = \frac{C^{\hat \phi^{\rm it} \phi^{\rm in}}_L}{C^{ \phi^{\rm in}  \phi^{\rm in}}_L}
\end{equation}

We see that when neglecting the delensed noise mean-field in the iterations, the effective normalization is decreased. For Simons-observatory, the effective normalization is decreased by about 15\% for $L<500$, and for CMB-S4 by about $20\%$. This is more important than the $\sim 3\%$ deviation we found on the polarization only estimator in \cite{Legrand:2021qdu}. 

The expression of the normalization as a Wiener filter is an approximation of the true response of the MAP estimator. This response is given by
\begin{equation}
	\label{eq:trueWF}
	\mathcal{W} = \left[\frac{1}{\cppfid} + \mathcal{H}\right]^{-1} \mathcal{R} \; ,
\end{equation}
where $\mathcal{R} = \frac{\delta g^{\rm QD}}{\delta X^{\rm dat}}\frac{\delta X^{\rm dat}}{\delta \phi}$ is the response of the quadratic gradient to the true lensing potential, analog of the standard QE response, and $\mathcal{H} = \frac{\delta^2 \ln \mathcal{L}}{\delta \phi^\dag \delta \phi} $ is the curvature of the likelihood estimated in $\phi^{\rm MAP}$. In the signal dominated regime we can approximate $\mathcal{H} \simeq \mathcal{R}$, which gives back the expression of the Wiener filter in Eq.~\eqref{eq:WF}. 

When ignoring the delensed noise mean-field in the likelihood, we can approximate the likelihood curvature as $\mathcal{H} \simeq \mathcal{R} - \mathcal{R}^{\rm MF}$, where we have subtracted the delensed noise mean-field response $\mathcal{R}^{\rm MF}$. This gives the following fiducial expression for the normalization
\begin{equation}
	\label{eq:wf_nomf}
	\mathcal{W}_L = \frac{\cppfid}{\cppfid + N_L^{(0)} - \frac{\mathcal{R}^{\rm MF}}{\mathcal{R}_L} \cppfid}\, .
\end{equation}
This prediction is shown as the dashed black lines in Figure~\ref{fig:mf_so_wf}. We see that in all experimental configurations, this prediction is a good approximation of the effective normalization obtained when neglecting the delensed noise mean-field in the iterations. The prediction for the response without mean-field is accurate at $\sim 8 \%$ for CMB-S4 at $L=500$.

\subsection{Delensing}
\label{sec:delensing}

Delensing the polarization B modes allows to put tight constraints on the tensor to scalar ratio. 
The delensing can be performed using a template of the lensed B modes, obtained by combining an estimate of the lensing potential with the observed E modes of polarization \cite{BaleatoLizancos:2020jby}. 
Here we assume that the lensing potential is reconstructed from the MAP temperature only estimator. In reality, one would use as well the polarization MAP estimator, or both temperature and polarization together. But we now wish to test the impact of the delensed noise mean-field on the obtained B mode template, which is better seen in the temperature only estimator. 

We construct the B mode template perturbatively, using the MAP lensing potential (obtained after 10 iterations), and the observed (lensed) E modes, which have been Wiener filtered to remove the noisy small scale modes. We do not re-normalize the lensing potential $\phi^{\rm it}$ to construct the lensing template. Indeed, we expect the MAP lensing potential to be optimally Wiener filtered.

\begin{figure}
	\centering
	\includegraphics[width=\columnwidth]{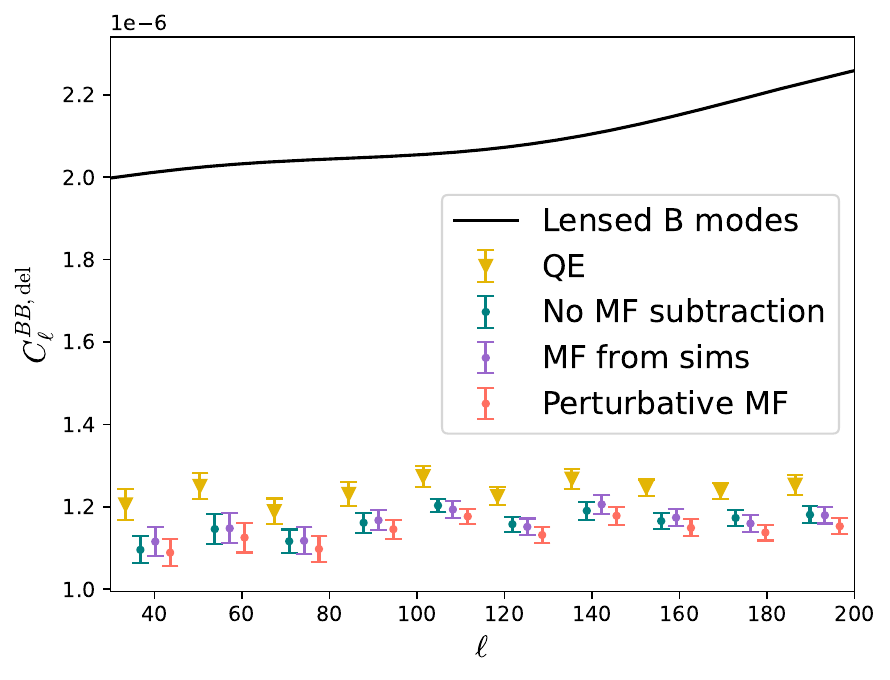}
	\caption{Residual $B$-modes power spectra after subtracting $B$-modes templates estimated from temperature only estimators, for a CMB-S4 configuration. The golden triangles show the delensing when the lensing template is obtained with a quadratic estimator. We show different scenarios for the MAP search: neglecting the mean-field (teal), mean-field estimated from Monte-Carlo simulations (purple), or with a perturbative prediction (pink). We binned the spectra in 10 multipole bins between $\ell_{\rm min}=30$ and $\ell_{\rm max} =200$, the bin centers are offset for clarity. The black line show the lensed B modes power spectrum for reference. }
	\label{fig:clbb}
\end{figure}

We show in Figure~\ref{fig:clbb} the residual B modes power spectrum when delensing with the lensing template $B^{\rm del} = B^{\rm dat} - \hat B^{\rm temp}$. We see that neglecting the delensed noise mean-field in the MAP iterations does not impact the delensing of the B modes. The efficiency of the delensing is not improved when considering the mean-field estimated from simulations or perturbatively at each iteration. 
The negligible impact of the delensed noise mean-field on delensing might be attributed to the fact that the lensed B modes are predominantly generated by the small-scale lensing potential, as illustrated in Figure 11 of \cite{Planck:2018lbu} and Figure 1 of \cite{Belkner:2023duz}. In contrast, the delensed noise mean-field primarily affects the large scales of the lensing potential. Furthermore, as shown in Figure~\ref{fig:corr_ratio}, neglecting the delensed noise mean-field still results in an optimal reconstruction of the lensing potential field.

\section{Conclusion}
\label{sec:conclusion}

Delensing the CMB removes anisotropies from the CMB signal but also creates new anisotropies in the noise maps. For any quadratic estimator applied on delensed CMB maps, these anisotropies will appear as a \textit{delensed noise mean-field}. This effect can, in principle, bias the field of interest reconstructed from these delensed CMB maps.

We have shown that the main effect is a modulation of the noise variance maps, more relevant in temperature than in polarization, and that to leading order this delensed noise mean-field traces very directly the large-scale lensing convergence map $\kappa$. We derived an analytic expression for the response of this delensed noise mean-field at first order in $\kappa$, with the help of calculations standard in the context of quadratic estimators.

We then studied the impact of this delensed noise mean-field within the framework of the maximum a posteriori (MAP) estimator and found that its noticeable effect is to change the normalization of the reconstructed lensing field.

Since at perturbative order the delensed noise mean-field is proportional to the lensing convergence map $\kappa$, it can be absorbed into the Gaussian prior of the MAP estimator. 
This also explains why neglecting it mostly impacts the normalization of the MAP estimator, since the Gaussian prior is driving the Wiener filter normalization.
If the mean-field response is indeed isotropic and perturbative, then the corresponding curvature matrix of the likelihood is effectively constant and diagonal. This means it would not induce mode mixing, but merely rescales the normalization of the reconstructed field.

In Eq.~\eqref{eq:wf_nomf} we have obtained a good analytical estimate of the MAP normalization in the case that the mean-field is neglected during the lensing reconstruction.
A more complete understanding of the normalization would require additional  analytical insight into the dense curvature matrix of the likelihood. This remains a formidable task.

Provided that the normalization is consistently estimated with a set of simulations, it is safe to neglect the delensed noise mean-field in the MAP search. Indeed, in particular for cross-correlation analysis \cite{Benoit-Levy:2013uxb, Carron:2022edh, ACT:2023oei}, one should be careful that whenever the noise properties of the survey varies across the sky, the normalization has to be recomputed for the footprint of the survey used in the analysis.
However we found that the delensed noise mean-field does not worsen the correlation coefficient of the reconstructed lensing field, and it does not impact the B modes delensing. 
Neglecting the delensed noise mean-field allows for faster MAP search, as it saves the computational time needed to estimate the mean-field in between each iterations. 

Therefore, for most cosmological analysis there appears to be little impact of the delensed noise mean-field, provided the change in normalization of the estimator is correctly accounted for. However, failure to do so in cross-correlations of CMB lensing with large-scale structure could lead to significant biases, of about 15\% to 20\% for a Simons-Observatory-like or CMB-S4-like experiment respectively. 

CMB foregrounds such as SZ, CIB and radio sources are also a source of anisotropy in the CMB maps, and they are highly non-Gaussian. They can bias the CMB lensing reconstruction, and since they are correlated to the density field, they could in principle be correlated with the delensed noise mean-field. Mitigations technique such as the bias hardening \cite{Namikawa:2012pe,Darwish:2025fip} can be used to reduce these foreground biases. However, a more complete study of the interplay of foregrounds and the delensed noise mean-field is left for future work.

Finally, our analysis showed that the delensed noise mean-field should not be neglected when applying a quadratic estimator sensitive to modulation-alike signals on delensed CMB maps, such as in searches for patchy reionization \cite{Dvorkin:2008tf, Darwish:2025fip}.

\section*{Acknowledgments}
LL acknowledges support from a SNSF Postdoc.Mobility fellowship (Grant No. P500PT\_21795).
JC acknowledges support from a SNSF Eccellenza Professorial Fellowship (No. 186879).
This work was supported by a grant from the Swiss National Supercomputing Centre (CSCS) under project ID sm80.
Some computations used resources provided through the STFC DiRAC Cosmos Consortium and hosted at the Cambridge Service for Data Driven Discovery (CSD3).

\appendix

\section{Quadratic estimators noise mean-fields}\label{app:MFresp}
\newcommand{\Wa}[0]{\ensuremath{\mathcal{W}_{\va}}}
\renewcommand{\D}[0]{\ensuremath{\mathcal D}}
\newcommand{\vb}[0]{ {\ensuremath{\boldsymbol{\beta} } }}
\newcommand{\lmax}[0]{ {\ensuremath{\ell_{\text{max}}} } }

In this section we discuss further the leading order full-sky MAP mean-field, its connection to more standard noise mean-fields, and how it can be computed fairly simply with already available lensing codes.

According to the arguments in the main text Sec.~\ref{sec:margMAP}, we can obtain the leading contribution from a quadratic estimator. This quadratic estimator is obtained in the standard manner, with the understanding that the filtering is anisotropic ($\va$ is present in the filtering matrix, since this is known source of anisotropy in the lensing reconstruction process beyond the quadratic estimator). 

By definition, mean-field calculations must be performed under the assumption that the input maps match precisely the likelihood model. Using this, we get the very compact formula
\begin{align}\label{eq:gMFMAP} \nonumber
	&\av{_{\pm1}\hat g^{\rm QE, del}_{\va}(\hn)} \\ &= -\sum_{l_1  m_1 l_2 m_2 }  \Wa^{l_1m_1l_2m_2}  \: \eth^{\pm}{}_{0}Y_{l_1m_1}(\hn)\:_{0}Y^*_{l_2m_2}(\hn)
\end{align}
where $ g^{\rm QE, del}_{\va}$ is introduced in Eq.\ref{eq:gdel} as the standard quadratic estimator on delensed maps, with anisotropic filtering depending on the deflection field $\va$, $\Wa$ is the Wiener-filtering matrix, $\Wa = (C^{-1} + \Nai)^{-1} \Nai$,  which may also be written formally as $C (C + \Na)^{-1}$, and $Y_{lm}(\hn)$ are the spherical harmonics.
This equation may look a bit mysterious but can be given some interpretation: consider first introducing two different locations $\hn_1$ and $\hn_2$ in the respective harmonics (instead of the identical $\hn$). 
After doing so, the object on the right-hand side in this equation can be seen as a matrix with indices $\hn_1, \hn_2$. This matrix represents the real-space operation of producing the most probable (in the rigorous, Bayesian sense -- this is what a Wiener-filter is doing) estimate of the gradient of a map at $\hn_1$, from the input map at $\hat n_2$, and the mean-field is the diagonal ($\hn_1=\hn_2 = \hn$) of this map-to-gradient matrix.
This comes about because the defining feature of (perturbative) lensing is  to introduce in the temperature map at any location $\hn$ a term proportional to its gradient at this very same location. 
In the absence of other anisotropies (in our full-sky case, evaluating the likelihood model at vanishing $\va$), no other mechanism does that. 
The value of the map and its gradient at identical locations are statistically independent: knowledge of the map at $\hn$ does not help estimating the gradient there, so that this average vanishes. 
If there are known anisotropies in the map (known in the sense of being implemented in the likelihood model), optimal recovery of the gradient may in general involve to some extent the map at the same location. 
Hence, the map and its recovered gradient are now related, but this is not lensing and must be subtracted.

It is useful to compare this formula to that obtained in the case of more standard quadratic estimators. 
To do this, to avoid confusion with lensing-induced anisotropies, let $\vb$ denote now any source of anisotropy. 
Quadratic estimators are very flexible and used in different variations, but the relevant aspects can be summarized as a two steps procedure: 1) inverse-covariance weighting of CMB maps using some covariance model 2) application of weights specific to the targeted source of anisotropy to these maps. 
Hence it is also useful to distinguish explicitly the true covariance matrix of the CMB in the presence of the anisotropy (let us call it $(C^{\rm dat})_{\vb}$) from the covariance matrix model used in step 1).
Let us call the filtering matrix $(F_{\vb})$. For optimal filtering, this matrix is the inverse of the data covariance, but in many cases the anisotropy is neglected in the filters, in which case the filtering $F_{\vb}\equiv F_0 $ is isotropic in harmonic space. For lensing, step 2) is what produces the Wiener-filtered gradient from the inverse-covariance filtered maps. 
This involves using a set of isotropic fiducial spectra, which we call $C_0^{\rm w}$, to possibly distinguish it from the CMB spectra. With this notation in place, the general lensing-QE response to $\vb$ is given by the same formula Eq.~\eqref{eq:gMFMAP}, but replacing the matrix $\Wa$ with
\begin{equation}\label{eq:generalWa}
	\Wa \rightarrow C_0^{\rm w} F_\vb C^{\rm dat}_\vb F_{\vb}.
\end{equation}
We can distinguish at least three cases of interest.
\begin{enumerate}
	\item The anisotropy lies in the CMB signal, and the QE uses isotropic filters. In this case, \eqref{eq:generalWa} becomes
	\begin{equation}
		C_0^{\rm w} F_0 C_{(1), \vb} F_0
	\end{equation}
	where $C_{(1),\vb}$ is the leading order change to the CMB spectra (a dense matrix).
	\item The anisotropy is now present not in the signal but in the noise CMB maps, but just as for point 1) it is neglected in the filters.  In this case the relevant matrix is
\begin{equation}
	C_0^{\rm w} F_0 N_{(1), \vb} F_0
\end{equation}
	I.e., this just amounts to trivially replace one set of CMB spectra by the noise spectra.

	\item The anisotropy is present in the noise CMB map, \emph{and} is implemented in the filters $F$. For weak anisotropies, the effect of anisotropic filtering is that the response changes sign: explicitly, to leading perturbative order, $F_\vb \sim F_0 - F_0 N_{(1), \vb} F_0$, and plugging this in \eqref{eq:generalWa} one gets	
	\begin{equation}\label{eq:WmatMAP}
		-C_0^{\rm w} F_0 N_{(1), \beta} F_0.
	\end{equation}
\end{enumerate}
Hence, comparing point 1) and 3) demonstrates very explicitly a recipe to compute the leading MAP mean-field response to $\va$ as follows: just use a standard lensing response calculation code, but replacing one set of CMB spectra in the numerator (the one entering the sky response to lensing $C_{(1), \va}$) by the noise spectra. The other spectra $C_0^{\rm w}$ and the filtering spectra remain unchanged. There is no explicit change in sign, since for the MAP mean-field in Eq.~\eqref{eq:WmatMAP} the anisotropy $\vb$ is tracing $-\va$, cancelling the sign in front, whereas the standard lensing response \eqref{eq:generalWa} has $\vb = + \va$.

\bibliography{biblio}

\end{document}